# Towards a general field equation for galaxies and galaxy clusters

A. Raymond Penner[1,2]

[1]Department of Physics, Engineering & Astronomy, Vancouver Island University, 900 Fifth St., Nanaimo, BC, Canada, V9R5S5
e-mail: raymond.penner@viu.ca
[2]Correspondence address: 1544 Scarlet Hill Rd., Nanaimo, BC, Canada, V9T1J3

**ABSTRACT**

The MONDian theory of AQUAL (AQUAdratic Lagrangian) and the theory of GRAS (GRavitational Anti-Screening) are alternatives to the theory of dark matter. When these theories are applied to galaxy dynamics they are in excellent agreement with observations including the galactic RAR (Radial Acceleration Relationship). However, when applied to galaxy clusters they do not explain the bulk of the missing mass. This manuscript develops a modified version of the GRAS/AQUAL field equation that can be extended to galaxy clusters. It involves just a single free parameter. The new field equation is then applied to a sample of galaxy clusters and checked against modeled galaxies and solar system constraints. Further to this, the modified field equation leads to an understanding of the difference between the galactic RAR and the RAR recently found for clusters.

**Key words:** gravitation – galaxies: general – galaxies: clusters: general – galaxies: clusters: intracluster medium – cosmology: dark matter

## 1. INTRODUCTION

Dark matter is an integral component of ΛCDM, the current standard model of cosmology (Aghanim et al. 2020; Blanchard et al. 2022; Trimble 1987; Bertone & Hooper 2018). The ΛCDM model has been successful in modeling a wide range of astronomical observations and provides the most accepted explanation for gravitational fields, greater than that provided by the baryonic mass, which are required to explain the dynamics of galaxies and galaxy clusters. This theory postulates that non-baryonic matter exists within galaxies and clusters that only interacts gravitationally or via the weak force. A particular success of the theory has been that the expected distribution of dark matter matches well with what is required to explain the hydrostatic stability of hot X-ray emitting intracluster gas in clusters. Indeed, it is the observations of galaxy clusters that initially led to the hypothesis that some sort of dark matter was required (Zwicky 1933).

Although successful with galaxy clusters, the theory of dark matter has had difficulty dealing with observations on the scale of galaxies. This includes the speed of galactic bars in disc galaxies (Roshan et al. 2021), the distribution of satellite galaxies (Pawlowski 2021), and the prediction that some types of dwarf galaxies should not contain dark matter (Kroupa 2014). A summary of such issues can be found in Banik & Zhao (2022). The most conspicuous observation that the theory of dark matter has difficulty with is the rotational velocity curves of spiral galaxies. These velocity curves are found to flatten out and approach a constant value in their outer regions (Famaey & McGaugh 2012). Weak gravitational lensing observations have shown that there is no apparent cut-off to this observation, with velocity curves remaining flat for hundreds of kiloparsecs, possibly out to 1 Mpc (Mistele et al 2024). Going beyond just the rotational curves flattening out, it has been found that there is a direct correlation between a galaxy's baryonic mass and this constant outer rotational speed. This relationship is known as the Baryonic Tully Fisher Relationship (BTFR) (McGaugh et al. 2000; McGaugh 2012) and is given by

$$M_{bar} = A_{BTFR} v_f^4 \tag{1a}$$

where $v_f$ is the velocity in the outer flattened section of a given spiral galaxy's rotation curve, $M_{bar}$ is the total baryonic mass of the galaxy, and $A_{BTFR}$ is a fitted parameter. The value of $A_{BTFR}$ as given by McGaugh (2012) is

$$A_{BTFR} = (47 \pm 6) M_\odot \text{ km}^{-4} \text{s}^4. \tag{1b}$$

The modeling of the expected dark matter distribution leads to a rotational profile (Navarro, Frenk & White 1996,1997) that does not naturally lead to this result (Wu & Kroupa 2015; Desmond 2017a, 2017b).

Recently the relationship between a galaxy's baryonic mass and its rotational speed has further been expanded to consider the complete rotational curve. This is highlighted by the work of McGaugh et al (2016) and Lelli et al (2017) who found a relationship between the radial acceleration as determined from the observed rotational curves and the predicted radial acceleration due to Newtonian gravity determined from the baryonic mass distribution of the galaxies. The following Radial Acceleration Relationship (RAR) was found to provide an exceptionally good fit to the data:

$$g = \frac{g_{bar}}{1 - e^{-\sqrt{g_{bar}/g_o}}}, \tag{2}$$

where g is the true acceleration as determined from the observed rotational curve, $g_{bar}$ is the Newtonian gravitational field as determined from the baryonic mass distribution and $g_o$ is a fitted gravitational parameter. From McGaugh et al's (2016) data set and their adopted stellar mass-to-light ratio, the fitted value of this observational parameter was found to be

$$g_o = (1.20 \pm 0.02) \times 10^{-10} \text{ms}^{-2}. \tag{3}$$

Analysing a set of disk galaxies from the SPARC database, Rodrigues et al (2018a, 2018b, 2020) have argued that the RAR and the constant $g_o$ is an emergent phenomenon arising from the averaging of results from individual galaxies. Using the same SPARC data set, this has been refuted by McGaugh et al (2018) and Kroupa et al (2018). For this manuscript it will be taken that the RAR, as given by Eq. (2), with the constant $g_o$, as given by Eq. (3), does hold for galaxies.

In the limit where $g_{bar} \ll g_o$, Eq. (2) leads to

$$g = (g_{bar} g_o)^{1/2}. \tag{4}$$

If the baryonic mass of the galaxy is spherically symmetric, $g_{bar}$ is given by

$$g_{bar} = \frac{GM_{bar}}{r^2} \tag{5}$$

where $M_{bar}$ is the total baryonic mass within r from the galactic centre. For a disc dominated galaxy, the value for $g_{bar}$ in Eq. (5) must take into account that the gravitational field within the disc of the galaxy is greater than for the case of spherical symmetry. Correcting for this leads to the Newtonian field within the disc being given by

$$g_{bar} = \alpha \frac{GM_{bar}}{r^2}, \tag{6}$$

where $\alpha = 1.32$ (McGaugh&Blok 1998). Equations (4) and (6) then lead directly to the BTFR as given by (1).

Although proponents of dark matter argue that the BTFR and the galactic RAR may be the result of the interplay between dark matter and baryons (Ludlow et al 2017, Navarro et al 2017, Famaey et al 2018), the BTFR and the galactic RAR have been the primary motivation behind alternative theories. Both the BTFR and the RAR strongly suggest that the baryonic mass of a galaxy is ultimately responsible for the total gravitational force that it experiences. This leads to two general possibilities.

The first possibility is that baryonic mass is solely responsible for the additional gravitational field. A modification to Newtonian theory, and therefore general relativity, is required to account for the observational properties of galaxies. MOdified Newtonian Dynamics or MOND falls under this category (Milgrom 1983a, 1983b, 1983c; Berkenstein & Milgrom 1984; Famaey & McGaugh 2012; Sanders 2014; Banik & Zhao 2022). MOND was originally proposed to explain galaxy rotation curves and the Tully-Fisher relation (Tully & Fisher 1977), the precursor to the BTFR. The two primary versions of MOND, AQUAL (AQUADratic Lagrangian) and QUMOND (QUasilinear MOND), both lead to a field equation that reduces in the spherically symmetric case and for $g \ll g_o$ to the BTFR. AQUAL will be the specific version of MOND considered in this manuscript.

The second possibility is that baryonic mass is indirectly responsible for the additional gravitational fields. Blanchet (Blanchet 2007a, 2007b, Blanchet & Le Tiec 2008) proposed the existence of a cosmic medium consisting of gravitational dipoles which become polarized in a gravitational field. The polarized dipole medium provides an additional contribution to the gravitational field surrounding a given mass.

Blanchet's gravitational dipoles are modeled as consisting of a pair of particles, one of positive gravitational mass, and the other of negative gravitational mass, with both particles having positive inertial mass. These dipoles will therefore align in a gravitational field. An unspecified internal force between the pair of particles that constitute the dipole is required to bind the dipole together. This force is also required to balance the local gravitational field so that the dipole distribution about a given mass is static. The internal force, and therefore the dipole moment, is modeled as either being dependent on the gravitational field (Blanchet 2007a, 2007b) or on the polarization field (Blanchet & Le Tiec 2008). This dependence is such that the model leads to the BTFR.

A variation of Blanchet's dipole theory is GRAS (GRavitational Anti-Screening) as proposed by Penner (2011, 2016a, 2016b, 2018, 2022). A similar theory has also been proposed by Hajdukovic (2011a, 2011b, 2020), but in this case it is not in agreement with solar system constraints (Banik & Koupa 2020). In GRAS it is hypothesised that the gravitational field surrounding a baryonic mass induces mass dipole moments in the quantum fluctuations in the surrounding vacuum. Unlike with Blanchet's dipoles, there are no unknown internal forces and the temporal nature of the fluctuations keep the dipole distribution static about a given mass.

The sea of virtual mass dipoles is analogous to the case in quantum electrodynamic theory (QED). In QED, the induced virtual electric dipoles in the vacuum results in a screening effect that leads to the observed charge of a particle being less than its actual bare charge. The vacuum behaves like a dielectric. In GRAS, the alignment of virtual mass dipoles results in an anti-screening effect that leads to the observed gravitational mass of a body being greater than its baryonic mass. The dependence that the resulting mass dipole moment density has on the gravitational field is modeled such that it leads to the BTFR and the galactic RAR. GRAS has also been applied to galactic rotation curves (Penner 2013, 2016a), binary galaxies (Penner 2023), and the Solar System (Penner 2020a, 2020b).

Gravitational dipole theories differ from MOND in that no modification of Newtonian theory is required. The contribution that a field of mass dipoles makes to the gravitational field is equal to a real mass density. However, the requirement of a particle or any entity with a positive inertia mass and a negative gravitational mass still poses a theoretical problem for any mass dipole theory.

Both AQUAL and GRAS lead to the same field equation (Penner 2022) and both are in excellent agreement with the BTFR and the RAR. However, while these theories are in good agreement with observations on the galactic scale, there are problems when applied to galaxy clusters (Sanders 2003, McGaugh 2015). Extrapolating MOND to clusters of galaxies has led to gravitational fields that are too weak to explain the hydrostatic equilibrium of the X-ray emitting cluster gas. More recently, ultra diffuse galaxies (UDGs) have been discovered within clusters (Mihos et al 2015, Koda et al 2015). Due to the relatively large gravitational fields of the clusters, MOND predicts that the UDG's should behave close to Newtonian and thereby suffer tidal disruptions (Bilek et al 2019, Hodson & Zhao 2017a). However, these UDG's typically do not show any signs of tidal disruptions. Therefore, these galaxies also appear to need significantly stronger gravitational fields than that currently provided by MOND. A unique problem is found with the Bullet cluster, where the galaxies have been spatially separated from the majority of the intracluster gas. For this special cluster, the gravitational fields provided by MOND also appear to be far too weak to explain gravitational lensing results (Angus et al 2007; Knebe et al 2009; Banik & Zhao 2022). All three of these problems that MOND has, would seem to indicate that diffuse baryonic mass distributions require significantly more gravity than what is apparently provided by the theory. Given that AQUAL and GRAS have the same field equation, these problems will also apply to GRAS.

To add another piece to the puzzle, galaxy clusters have recently been found to have their own RAR (Tian et al 2020, 2024). Analysis of 20 high-mass clusters targeted by the Cluster Lensing And Supernove survey with Hubble (CLASH) leads to the following relationship for clusters (Tian et al 2020)

$$\ln(g) = 0.51^{+0.04}_{-0.05} \ln(g_{bar}) - 9.80^{+1.07}_{-1.08}. \tag{7}$$

Fixing the slope at m=0.50, Eq. (7) can be rewritten as

$$g = (g_{bar} g_{oc})^{1/2} \tag{8}$$

with $g_{oc}$, a fitted gravitational parameter for clusters, given by

$$g_{oc} = (2.02 \pm 0.11) \times 10^{-9} \text{ms}^{-2}. \tag{9}$$

Both the galactic RAR, Eq. (2), and the new cluster RAR, Eq. (7) are included in Figure 7. As is seen in the figure, the slope for both relationships with $g_{bar} \ll g_o$ for galaxies and with $g_{bar} \ll g_{oc}$ for clusters is similar, leading respectively to Eqs. (4) and (8). However, given that the same $g_{bar}$ leads to two vastly different total gravitational fields indicates that there can be no universal RAR that just depends on $g_{bar}$. As will be seen, AQUAL and GRAS similarly have the total gravitational field dependent just on $g_{bar}$. This explains why they also fail when they are extended to galaxy clusters.

If the stronger gravitational fields found with galaxies and clusters are not due to dark matter but to current gravitational theory being incomplete, the ultimate goal is to derive a theoretically based general field equation that encompasses all. The goal of this particular manuscript is less ambitious. The goal here is to take a step towards the final gravitational field equation by modifying the GRAS/AQUAL field equation so that it applies not only to galaxies but also to galaxy clusters. Although there are other significant problems that GRAS and AQUAL have, the fact that when applied to clusters a large mass discrepancy still exists may be the most serious. This modified GRAS/AQUAL equation will provide some insights into the form of the true field equation. In addition, it will lead to an understanding of the difference between the galactic RAR and the cluster RAR and how these two RAR's are compatible with each other.

The outline of this manuscript is as follows: Section 2 will discuss the cluster sample and galaxy model that will be used in the analysis. Section 3 will discuss the current field equations of GRAS and AQUAL followed by the proposed modified field equation. Section 4 will then apply the old and new field equations to the cluster sample and the galaxy model. Section 5 will provide a conclusion and a discussion of the results. Included in Section 5 will be a proposed method for determining whether the additional gravitational fields required for galaxies and clusters is due to dark matter or if a modification to current gravitational theory is needed.

## 2. CLUSTER SAMPLE AND GALAXY MODEL

The GRAS/AQUAL field equations, as well as the galactic and cluster RARs, require as input $g_{bar}$, the gravitational field due to the baryonic mass. To determine $g_{bar}$, the baryonic density profile, $\rho_{bar}$, of a galaxy or a galaxy cluster first needs to be determined. Once $\rho_{bar}$ is found, the enclosed baryonic mass within r is given by

$$M_{bar}(r) = \int_0^r 4\pi r^2 \rho_{bar}(r) dr. \qquad (10)$$

and $g_{bar}$ can then be determined by Eq. (5).

### 2.1 Galaxy cluster sample

For this manuscript, the same 12 clusters that Hodson and Zhao (2017b) used in their analysis, selected from the Chandra galaxy cluster sample (Vikhlinin et al 2006), will be considered. These 12 clusters are all nearby (z<1) and are in the baryonic mass range $10^{13}$ to $10^{15}$ $M_\odot$. The 12 clusters are listed in Table 1 along with their inner and outer radii that will be used in this manuscript. The inner radii are as given by Vikhlinin et al (2006), were they excluded the central temperature bin. The outer radii are set equal to $1.5 r_{500}$, as per Hodson and Zhao (2017b), where the $r_{500}$ values are as given by Vikhlinin et al (2006).

The dominant baryonic mass component of these clusters is the X-ray emitting gas. To model the density of this gas, Vikhlinin et al (2006) chose a three-term model leading to the following expression for $\rho_{gas}(r)$,

$$\rho_{gas}(r) = 1.252 m_p\, n_o \left( \frac{(r/r_c)^{-\alpha}}{(1+(r/r_c)^2)^{3\beta-\alpha/2}} \frac{1}{(1+(r/r_s)^\gamma)^{\varepsilon/\gamma}} + \frac{1}{(1+(r/r_{c2})^2)^{3\beta_2}} \right)^{1/2} \qquad (11)$$

where $m_p$ is the mass of a proton and $n_o$ is the central number density. The value for $n_o$ and the values for the scale radii $r_c$, $r_s$, and $r_{c2}$, along with the dimensionless parameters $\alpha$, $\beta$, $\gamma$, $\varepsilon$, and $\beta_2$ are given for all 12 clusters in Vikhlinin et al (2006) as well as in Hodson and Zhao (2017b).

| Cluster | $r_{inner}$ (kpc) | $r_{outer}$ (kpc) |
|---|---|---|
| A133 | 40 | 1511 |
| **A478** | 30 | 2006 |
| **A1413** | 20 | 1949 |
| **RXJ1159** | 10 | 1050 |
| **MKW4** | 5 | 951 |
| **A907** | 40 | 1644 |
| **A262** | 10 | 975 |
| **A383** | 25 | 1416 |
| **A1795** | 40 | 1853 |
| **A1991** | 10 | 1098 |
| **A2029** | 20 | 2043 |
| **A2390** | 80 | 2124 |

**Table 1:** The 12 clusters along with their inner and outer radii that are considered in the manuscript.

After the gas, the only other significant baryonic mass component are the galaxies themselves. Following Hodson and Zhao (2017b) it will be taken that only the brightest cluster galaxy (BCG), located in the centre of the cluster, will have a significant contribution to the cluster's gravitational field. The other galaxies are taken to contribute a relatively small amount to the overall baryonic mass of the cluster and to the resulting gravitational field. For the density profile for the BGC, the following Hernquist model from Hodson and Zhao (2017b) is used

$$\rho_{BCG}(r) = \frac{M_{BCG} h}{2\pi r (r+h)^3} \qquad (12)$$

where h, the Hernquist scale length, is set equal to 30 kpc for all clusters, as per Hodson and Zhao. The mass, $M_{BCG}$, of the BCG, following from the work of Schmidt and Allen (2007), is given by

$$M_{BCG} = 5.3 \times 10^{11} \left(\frac{M_{500}}{3 \times 10^{14} M_\odot}\right)^{0.42} \qquad (13)$$

where $M_{500}$ is the enclosed mass at $r_{500}$, the radius at which the average density is 500 times the critical density of the universe. The values for $M_{500}$ are provided by Vikhlinin et al (2006). The baryonic density profile is then given by $\rho_{bar}(r) = \rho_{gas}(r) + \rho_{BCG}(r)$.

Figure 1 shows the resulting baryonic mass profiles, obtained from Eq. (10), for the 12

clusters. For this figure, the mass profiles have been extrapolated out to 10 Mpc. The 12 clusters have been split into two groups. In Figure 1a are the 6 clusters where the baryonic mass approaches a constant value at large radii. This is what is expected for bounded clusters. In Figure 1b are the other 6 clusters, which appear to be unbounded, with their baryonic mass still significantly increasing even beyond 10 Mpc. For these clusters it is difficult to set a cut-off radius. As will be discussed this is a problem. As such, the subsequent analysis will focus on the 6 bounded clusters of Figure 1a.

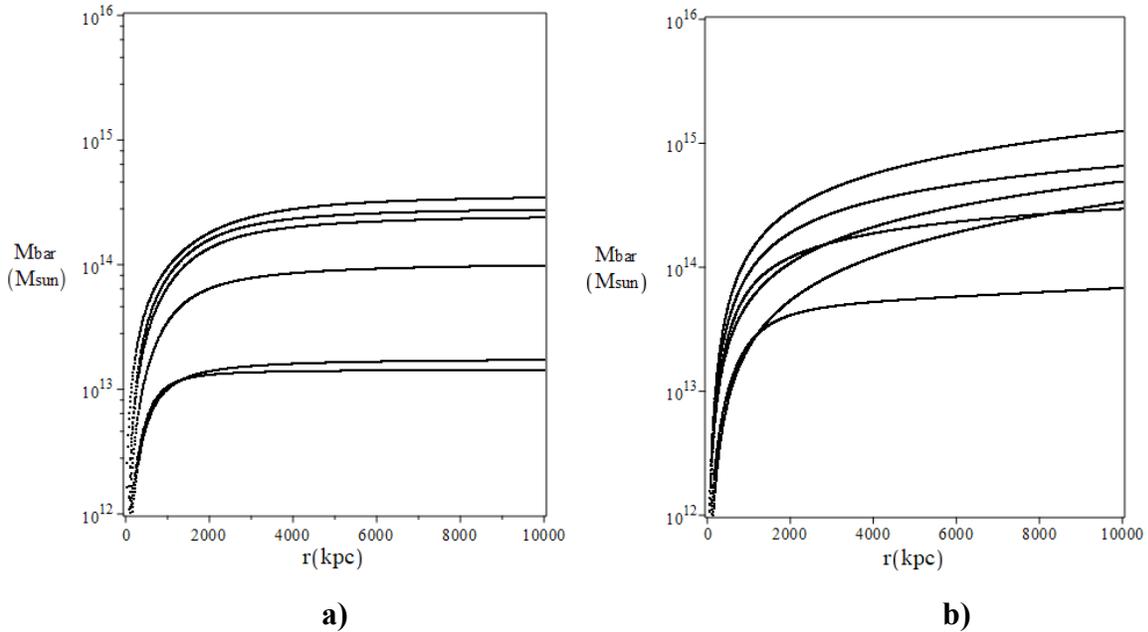

a)          b)

**Figure 1:** The baryonic mass profiles of the 12 clusters obtained from Eqs. (10), (11), and (12). In figure a) are the 6 clusters which behave as bounded clusters. These are the galaxy clusters A133, A478, A 1413, RXJ1159, MKW4, and A907. In figure b) are the other 6 clusters whose mass profiles based on Eqs. (10), (11), and (12), keep increasing.

Given the density and temperature profiles for a given cluster, one can also estimate $M_{dyn}(r)$, the total mass enclosed within the radius r, required to have the gas in hydrostatic equilibrium. The following expression for $M_{dyn}(r)$ is given in Vikhlinin et al (2006),

$$M_{dyn}(r) = -3.68 \times 10^{13} M_\odot T(r) r \left( \frac{d \ln \rho_{gas}}{d \ln r} + \frac{d \ln T}{d \ln r} \right) \quad (14)$$

where T is in units of keV and r is in units of Mpc. The required temperature profile of the gas, T(r), as given by Vikhlinin et al (2006), is

$$T(r) = T_0 \frac{(r/r_{cool}+T_{min}/T_0)}{(r/r_{cool})^{a_{cool}}+1} \frac{(r/r_t)^{-a}}{((r/r_t)^b+1)^{c/b}}. \tag{15}$$

The values for the gas parameters are given in Vikhlinin et al (2006).

In the theory of dark matter, the contribution of the dark matter is most often given by the following NFW profile

$$M_{NFW}(r) = 4\pi\rho_s r_s^3 \left[\ln\left(1+\frac{r}{r_s}\right) - \left(1+\frac{r_s}{r}\right)^{-1}\right] \tag{16}$$

where $r_s$ is the radius of the NFW profile and

$$\rho_s = \frac{500}{3}\left(\frac{r_{500}}{r_s}\right)^3 \frac{\rho_{crit}}{\log\left(1+\frac{r_{500}}{r_s}\right) - \left(1+\frac{r_s}{r_{500}}\right)^{-1}} \tag{17}$$

where $\rho_{crit}$ is the critical density. Values for $r_s$ and $r_{500}$ are given in Hodson and Zhao (2017b). Therefore, the mass profile for a cluster, according to the theory of dark matter, will be given by

$$M_{DM}(r) = M_{gas}(r) + M_{BCG}(r) + M_{NFW}(r). \tag{18}$$

As will be seen, the dark matter profile given by Eq. (18) is in good agreement with $M_{dyn}(r)$, determined from Eq. (14).

## 2.2 Spiral galaxy model

When modifying the GRAS/AQUAL field equation so that it can be extended to galaxy clusters, it is required that the new field equation still agrees with galaxy observations, specifically the velocity curves of spiral galaxies. In order to provide a check on this, a model of the baryonic mass distribution of a spiral galaxy is needed. For the stellar mass component, the following spherically symmetric models for the bulge, $\rho_{bulge}(r)$, and disc, $\rho_{disc}(r)$, stellar mass distributions were used,

$$\rho_{bulge}(r) = \frac{M_{bulge}}{2\pi}\frac{a}{r}\frac{1}{(r+a)^3} \tag{19}$$

and

$$\rho_{disc}(r) = \frac{M_{disc}}{8\pi h^3}e^{-r/h} \tag{20}$$

where $M_{bulge}$ and $M_{disc}$ are the total baryonic masses of the bulge and disc respectively, and a and h are scale radii.

For the purposes of this manuscript, the stellar mass distributions were determined corresponding to representative Hubble stage T=3 or Sb type galaxies with $M_{disc}/M_{bulge}$ = 3.22 and total stellar masses of $20 \times 10^9$ $M_\odot$, $60 \times 10^9$ $M_\odot$, and $180 \times 10^9$ $M_\odot$. Table 2 shows the parameters used for these modeled galaxies (Penner 2013).

| $M_{star}$ ($M_\odot$) | a (kpc) | h (kpc) |
|---|---|---|
| 180×10⁹ | 0.671 | 5.40 |
| 60×10⁹ | 0.472 | 3.80 |
| 20×10⁹ | 0.332 | 2.67 |

**Table 2:** The values of a (kpc), and h (kpc) used for the 3 modeled galaxies.

In addition to the stellar component, galaxies also contain significant amounts of gas (Ma et al 2025, Read & Trentham 2005). For the gas component of the above three model galaxies, the following baryonic mass distributions, as given by Ma et al (2025), of the H1 gas component, $\rho_{H1}(r)$, and the hot gas halo, $\rho_{hot}(r)$, were used,

$$\rho_{H1}(r) = \frac{\Sigma_0 K_0(r/R_s)}{\pi R_s} \qquad (21)$$

and

$$\rho_{hot}(r) = \rho_0[1 + (r/r_c)^2]^{-\frac{3}{2}\gamma}. \qquad (22)$$

where $K_0$ is the zeroth-order modified Bessel function of the second kind. The specific set of gas parameters that will be used are the values, as given by Ma et al (2025), corresponding to the mass bins that the above three modeled galaxies fall in. These values, from their subset of 129,278 star forming galaxies, are provided in Table 3.

Using Ma et al's gas model, along with the values given in Table 3, results in the gas fraction for the three modeled galaxies ranging from approximately 25% to 40%. The overall baryonic density profile for the galaxies is then given by $\rho_{bar}(r) = \rho_{bulge}(r) + \rho_{disc}(r) + \rho_{H1}(r) + \rho_{hot}(r)$ with the baryonic mass distribution then determined by Eq. (10).

| $M_{star}$ ($M_\odot$) | $\Sigma_0$ ($10^{10} M_\odot$ kpc⁻²) | $R_s$ (kpc) | $\rho_0$ ($10^5 M_\odot$ kpc⁻³) | $r_c$ (kpc) | $\gamma$ |
|---|---|---|---|---|---|

| 180×10⁹ | 2.18 | 0.32 | 0.04 | 103.79 | 1.62 |
| 60×10⁹ | 0.82 | 0.62 | 0.03 | 75.31 | 1.62 |
| 20×10⁹ | 0.37 | 0.35 | 0.02 | 61.54 | 1.61 |

**Table 3:** The values of $\Sigma_o$, $R_s$, $\rho_o$, $r_c$, and $\gamma$ that were used for the 3 modeled galaxies.

## 3. THEORY
### 3.1 AQUAL and GRAS field equations

In Newtonian gravitational theory, the field equation is given by

$$\boldsymbol{\nabla} \cdot \boldsymbol{\nabla} \Phi_{bar} = 4\pi G \rho_{bar}. \tag{23}$$

where $\Phi_{bar}$ is the Newtonian gravitational scalar potential and $\rho_{bar}$ is the baryonic mass density. The Newtonian gravitational field, $\mathbf{g_{bar}}$, in turn is given by

$$\mathbf{g_{bar}} = -\boldsymbol{\nabla} \Phi_{bar}. \tag{24}$$

In the AQUAL version of MOND, and MOND in general, Newtonian gravitational theory and GR are modified, but the gravitational source remains just $\rho_{bar}$. The nonrelativistic AQUAL field equation is given by

$$\boldsymbol{\nabla} \cdot (\mu(g_A/g_o)\boldsymbol{\nabla}\Phi_A) = 4\pi G \rho_{bar}, \tag{25}$$

with the resulting AQUAL gravitational field, $\mathbf{g_A}$, given by

$$\mathbf{g_A} = -\boldsymbol{\nabla}\Phi_A. \tag{26}$$

The function $\mu(g_A/g_o)$ is referred to as the interpolating function. In the case of spherical symmetry, the field equation given by Eq. (25) can be expressed by

$$\mu(g_A/g_o)g_A = g_{bar}. \tag{27}$$

The interpolating function $\mu(g_A/g_o)$ therefore determines the deviation between the AQUAL and Newtonian gravitational fields.

To have agreement with both Newtonian theory within the solar system (Hees et al 2014, Hees et al 2016, Aurelien et al 2016) and the BTFR, the interpolating function must be such that

$$\mu(g_A/g_o) \rightarrow 1 \quad \text{for } g_A \gg g_o, \tag{28a}$$

and

$$\mu(g_A/g_o) \rightarrow \frac{g_A}{g_o} \quad \text{for } g_A \ll g_o. \tag{28b}$$

For $g_A \ll g_o$, by Eqs. (27) and (28b) the relationship between the Newtonian gravitational field $g_{bar}$ and the AQUAL gravitational field $g_A$ leads to Eq. (4), which in turn leads to the BTFR.

Although in AQUAL the only gravitational source is the baryonic mass, for computational purposes, especially when dealing with a non-spherically symmetric mass distribution, a phantom dark matter density $\rho_{PDM}$ can be defined as

$$\rho_{PDM} = \frac{1}{4\pi G} \nabla \cdot \left((1 - \mu(g_A/g_o))\nabla \Phi_A\right). \tag{29}$$

The field equation can then also be expressed as

$$\nabla \cdot \nabla \Phi_A = 4\pi G(\rho_{bar} + \rho_{PDM}). \tag{30}$$

Knowing $\rho_{bar}$ and determining the distribution $\rho_{PDM}$, one can then use Newton's gravitational theory to determine $\mathbf{g_A}$.

In GRAS's gravitational theory, baryonic masses are taken to induce mass dipole moments in a surrounding sea of quantum fluctuations. This leads to an anti-screening of the baryonic mass and hence a larger observed mass. The dependence that the resulting mass dipole moment density, $\mathbf{P}$, has on the total gravitational field $\mathbf{g_G}$ is given by the theory to be

$$\mathbf{P} = \frac{1}{4\pi G} f(g_G/g_o)\mathbf{g_G}, \tag{31}$$

with the function $f(g_G/g_o)$ incorporating any nonlinearity between $\mathbf{P}$ and $\mathbf{g_G}$. Analogous to a dielectric in electromagnetism, the equivalent mass density $\rho_{dipole}$ corresponding to this field of mass dipoles will be given by

$$\rho_{dipole} = -\nabla \cdot \mathbf{P} \tag{32a}$$

$$= \frac{1}{4\pi G} \nabla \cdot (f(g_G/g_o)\nabla \Phi_G). \tag{32b}$$

In GRAS, Newtonian gravitational theory holds but $\rho_{dipole}$ must now be included as a gravitational source. As such the nonrelativistic GRAS field equation is given by

$$\nabla \cdot \nabla \Phi_G = 4\pi G(\rho_{bar} + \rho_{dipole}) \tag{33}$$

or by substituting from Eq. (32b)

$$\nabla \cdot \nabla \Phi_G = 4\pi G \rho_{bar} + \nabla \cdot (f(g_G/g_o)\nabla \Phi_G) \tag{34}$$

In the case of spherical symmetry, the field equation given by (34) becomes

$$g_G = g_{bar} + f(g_G/g_o)g_G \tag{35a}$$

or

Penner – Towards a general field equation$$(1 - f(g_G/g_o))g_G = g_{bar} \,. \tag{35b}$$

In order have agreement with both Newtonian theory within the solar system and the BTFR, the interpolating function in GRAS must be such that

$$f(g_G/g_o) \to 0 \qquad \text{for } g_G \gg g_o, \tag{36a}$$

and

$$f(g_G/g_o) \to 1 - \frac{g_G}{g_o} \quad \text{for } g_G \ll g_o. \tag{36b}$$

From Eqs. (25) and (34), it can be seen that the field equations for AQUAL and GRAS are equivalent, with the interpolating functions for the two related by

$$\mu(g_A/g_o) = 1 - f(g_G/g_o) \,. \tag{37}$$

The determined distributions for $\rho_{dipole}$ and $\rho_{PDM}$ and the generated gravitational fields, $g_G$ and $g_A$, will be identical. This common gravitational field that falls out of GRAS and AQUAL will be referred to as $g_{GA}$ for the remainder of the manuscript. The equivalence of both theories is beneficial as previous results obtained specifically using AQUAL or specifically using GRAS will therefore apply to both theories.

Although the two theories have the same field equation, conceptually they are quite different. In the case of AQUAL, $\rho_{PDM}$ is taken to be a phantom density, a tool used to assist in solving the field equation. In GRAS, $\rho_{dipole}$ is a real gravitational source, equivalent to a real $\rho_{bar}$. In some sense GRAS is a mixture of MOND and dark matter, with $\rho_{dipole}$ playing the role of dark matter.

In addition to having agreement with the BTFR and solar system observations, the interpolating function also needs to agree with the more general galactic RAR. The following interpolating function, found using GRAS (Penner 2023),

$$f(g_{GA}/g_o) = \left(1 + \frac{g_{GA}}{2g_o}\right)^{-2}, \tag{38}$$

satisfies Eq. (36) and leads to very good agreement with the RAR. As such, Eq. (38) will be the interpolating function used in this manuscript. In the case of AQUAL, the equivalent interpolating function would be

$$\mu(g_{GA}/g_o) = 1 - \left(1 + \frac{g_{GA}}{2g_o}\right)^{-2}. \tag{39}$$

A major complication with the GRAS/AQUAL field equations, Eqs. (27) and (35b), is that they are non-linear in $g_{GA}$. As such, in general, external gravitational fields will need to be included in calculations. This is referred to as the External Field Effect (EFE). Examples of how this is handled in MOND is provided by Milgrom (1986) and by Banik & Zhao (2018). In these cases, a point mass in a uniform external gravitational field is considered. Applying GRAS to the case of the Sun in the approximately uniform external gravitational field of the Galaxy is considered in Penner (2020a). For the purposes of this manuscript, it will be taken that the EFE is negligible and the galaxies and clusters considered are isolated.

If the baryonic mass is not spherically symmetric, then either Eqs. (25) or (34) needs to be used in order to calculate the resulting gravitational field. Examples of handling this in MOND is provided by Milgrom (1986) and De-Chang, Matsuo, and Starkman (2010). An example of handling this in GRAS is provided by Penner (2020a, 2023). In all these examples the baryonic mass distributions considered are axisymmetric.

For this manuscript, as any EFE is neglected and the baryonic mass distributions are spherically symmetric, Eqs. (27) and (35) apply. As such, the resulting gravitational fields only depend on $g_o$ and on $g_{bar}$, which in turn only depends on the enclosed baryonic mass. So as discussed in the introduction, in their present form, these field equations will fail when extended to galaxy clusters.

### 3.2 Modified field equations

There have been previous attempts to modify MONDian theories so that they can be extended to handle clusters. In any modifications to MONDian theories, it is required to preserve the theories' behaviour on galaxy scales, where they work so well, while significantly increasing the gravitational fields within clusters and UDGs. In general, this has involved going from just the single parameter, $a_o$ or $g_o$ (depending on the theory), to having 2 or more parameters. In different versions of EMOND (Zhao & Famaey 2012, Hodson & Zhao 2017b) the interpolating function is a function of the gravitational potential, in addition to the gravitational field, along with two fitted parameters. The gravitational potential is in general greater in magnitude within clusters than in galaxies and in EMOND this leads to a greater boost to the overall gravitational field. EMOND has also been applied to UDGs (Hodson & Zhao 2017a). Given that for both clusters and UDGs, EMOND leads to better agreement with observations does suggest that the cluster and UDG issues are one in the same. Refractive gravity (RG) is another alternative modification to gravitational theory that has been applied to both galaxies and clusters (Matsakos

& Diaferio 2016, Cesare et al 2020). In RG, the Poisson equation is modified by the introduction of the gravitational permittivity, a monotone increasing function of the local mass density. The equivalent interpolation function for this formulation has three fitted parameters.

Although such modifications to MONDian and other gravitational theories certainly lead to better results with regards to clusters, this is to be expected with the increased number of parameters. In this manuscript the GRAS/AQUAL field equation will be modified so that it applies to both galaxies and galaxy clusters, but will just have the one parameter, $g_o$, as is given by Eq. (3).

As a cornerstone of EMOND and RG, a crucial difference between galaxies and galaxy clusters is their distribution of the baryonic mass. This in turn affects the behaviour of the total gravitational field, or in the case of EMOND differences in the overall gravitational potential. Figure 3 shows how the total gravitational field depends on radial distance for the modeled galaxies and the 6 clusters of Section 2. For the galaxies, the gravitational field as given by Eq. (2) is used, as this has been found to be in good agreement with galactic velocity curves. Likewise, for the galaxy clusters, $g_{DM}$ calculated using Eq. (18) is used, as it has been found to be in good agreement with the determination of the dynamical mass distribution as given by (14). As is seen, the dependence that the gravitational field has on the radial distance for clusters is quite different to that for galaxies. For the given galaxy models, the fall off in g goes as approximately $r^{-1}$, while for the clusters, with their diffuse baryonic matter distribution, g initially falls off at a much slower rate.

To quantify the dependence that the gravitational field has on radial distance, the following parameter β is defined,

$$\beta = -\frac{g}{rg'}. \tag{40}$$

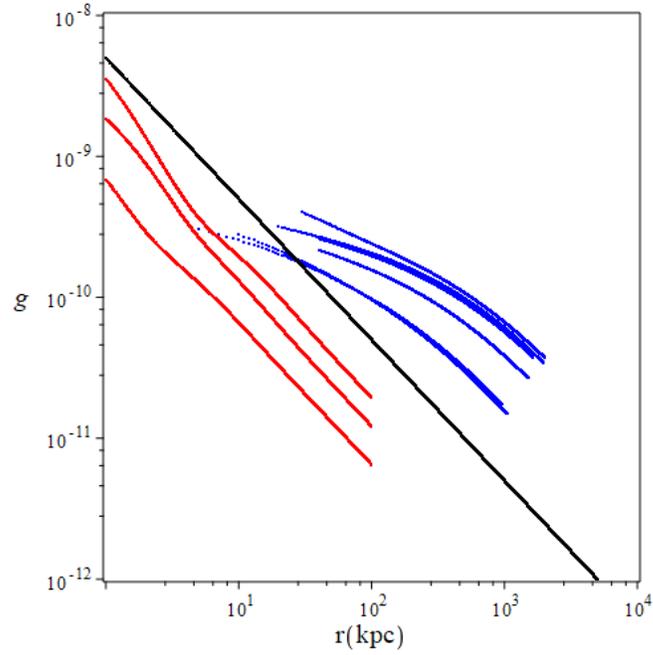

**Figure 2:** The gravitational field for the galaxy clusters using $g_{DM}$ (blue lines) and the model galaxies using g from Eq. (2) (red lines). The radial ranges for the clusters are given in Table 1. For the model galaxies the outer radial range has been set to 100 kpc. The black line corresponds to g proportional to $r^{-1}$.

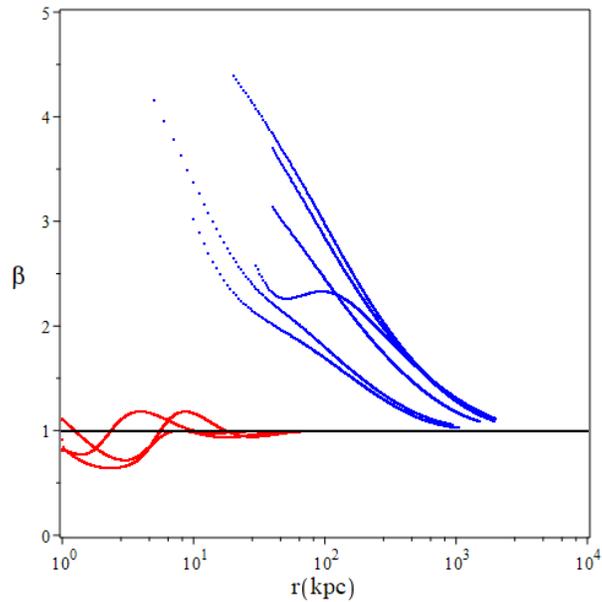

**Figure 3:** The value of β versus radial distance for the 6 clusters (blue lines) and the modeled galaxies (red lines).

For a gravitational field falling off as $r^{-1}$, as found with the modeled galaxies, $\beta = 1$, while in general, for a field falling off as $r^{-n}$, $\beta = n^{-1}$. Figure 3 shows $\beta$ versus r for the modeled galaxies and clusters. As is seen, the value of $\beta$ for the clusters for $r \lesssim 100$ kpc is much greater than that for galaxies.

Multiplying the interpolating function $f(g/g_o)$ as given by Eq. (38) by $\beta$ should therefore have a large impact on clusters but a much smaller impact on galaxies. The modified field equation in the case of spherical symmetry will then be given by

$$g_{GA} = g_{bar} + \beta f(g_{GA}/g_o) g_{GA} . \tag{41}$$

In the case of GRAS, this corresponds to the dependence of the mass dipole moment density being given by

$$\mathbf{P} = \frac{1}{4\pi G} \beta f(g_{GA}/g_o) \mathbf{g_{GA}}. \tag{42}$$

Substituting for $\beta$ and rearranging Eq. (41) results in

$$g_{GA}' = -\frac{g_{GA}}{r} \left(1 - \frac{g_{bar}}{g_{GA}}\right)^{-1} f(g_{GA}/g_o) . \tag{43}$$

Integrating Eq. (43) then leads to

$$g_{GA} = g_{GAo} + \int_r^{r_o} \left[\frac{g_{GA}}{r}\left(1 - \frac{g_{bar}}{g_{GA}}\right)^{-1} f(g_{GA}/g_o)\right] dr \tag{44}$$

were $g_{GAo}$ is the gravitational field at a given reference distance $r_o$.

As discussed in the previous section, the standard GRAS/AQUAL field equation just depends on $g_o$ and $g_{bar}$, and $g_{bar}$, in the case of spherical symmetry and neglecting any EFE, just depends on the total enclosed baryonic mass within the specified radial distance r. It does not matter what the baryonic mass distribution is outside of r, just what the total enclosed baryonic mass is. The field equation as given by Eq. (44) is quite different. The value of $g_{bar}$, and therefore the baryonic mass, outside of r matters. The complete baryonic mass distribution matters. This is a consequence of having the field equation as given by Eq. (43) involving both the gravitational field, $g_{GA}$, and its derivative, $g_{GA}'$.

To solve the field equation as given by Eq. (44) for a given $g_{bar}$ profile, the value of $g_{GAo}$ at a specified radius $r_o$ needs to be determined. The key to determining these values is that when at a distance well beyond the baryonic mass distribution of a given system, that system starts behaving as a point mass. At such distances the gravitational fields for both galaxies and clusters will behave identically, with $g_{bar}$ being given by Eq. (5) with $M_{bar}$ being the total baryonic mass of the system. In addition, we know what the gravitational field should be in this region. The galactic RAR given by Eq. (2) applies to both within as well as to far outside the baryonic mass distribution of a galaxy. Therefore, it will also apply to clusters when far outside of their baryonic mass distribution. For a radius of $r_o$, which is far outside of the cluster's baryonic mass distribution, $g_{GAo}$ will therefore be given by the galactic RAR,

$$g_{GAo}(r_o) = \frac{g_{bar}(r_o)}{1 - e^{-\sqrt{g_{bar}(r_o)/g_o}}}. \tag{45}$$

The technique that is used to determine what the value of $r_o$ should be for the clusters, is to consider the effect that $r_o$ has on the gravitational field calculated at the outer radiuses as listed in Table 1. The criterion is that the $r_o$ chosen should be such that the impact of doubling its value affects the gravitational field at the outer radiuses, $g_{GA}(r_{outer})$, by less than 1%. For the 6 clusters being considered this leads to a value for $r_o$ of 10 Mpc, the limit shown in Figure 1. For a $r_o$ of 20 Mpc the values of $g_{GA}(r_{outer})$ range from being 0.27% to 0.87% higher than for the selected value of 10 Mpc. For comparison for the 6 clusters of Figure 1b the gravitational fields at $r_{outer}$ range from being 8.0% to 24% higher as $r_o$ increases from 10 Mpc to 20 Mpc. To handle unbound clusters such as these, $r_o$ could be increased to much larger values so that $g_{GA}(r_{outer})$ eventually stays relatively consistent. Otherwise, another method would need to be used. One possibility is that observations of line-of-sight velocities and positions of the outer galaxies within a given cluster could be used to determine an estimate of $g_{GA}(r_o)$ at a given $r_o$. The GRAS/AQUAL field equation as given by Eq. (44) could then be used to determine the gravitational fields at other radii.

For the model galaxies, the criterion for choosing the value of $r_o$ was where the calculated gravitational field values at 50 kpc changed by less than 1% when the value of $r_o$ was doubled. This leads to a value of $r_o$ equal to 150 kpc. For a $r_o$ of 300 kpc the values of $g_{GA}(50$ kpc$)$ range from being 0.08% to 0.65% lower than for when $r_o$ was set to 150 kpc.

The field equation given by Eq. (41) with $\beta$ as given by Eq. (40) is valid only for the case of spherical symmetry. In the case of non-spherically symmetric baryonic mass distributions, the parameter $\beta$ can be expressed as

$$\beta = \frac{1}{2}\left(1 - \frac{\nabla \cdot \mathbf{g}}{g'}\right), \tag{46}$$

which in the case of spherical symmetry reduces to Eq. (40). For an axisymmetric distribution the value of the gravitational field at a specific position, ($\rho_o$, $z_o$) will still be needed. This will be more challenging and overall, the determination of the GRAS/AQUAL gravitational field will be much more complex.

## 4 RESULTS

### 4.1 Galaxy clusters

Substituting $r_o$ = 10 Mpc and $g_{GAo}$, as given by Eq. (45), into Eq. (44) allows for the gravitational field profile, $g_{GA}$, to be determined for each of the 6 clusters. From this the mass profile for each cluster is determined. The result is shown in Figure 4, along with the mass profile found using the original GRAS/AQUAL field equation (35). The ranges of radii used for each cluster are as given in Table 1. Included in the figures for each cluster are the mass profiles corresponding to $M_{bar}$, $M_{dyn}$ and $M_{DM}$.

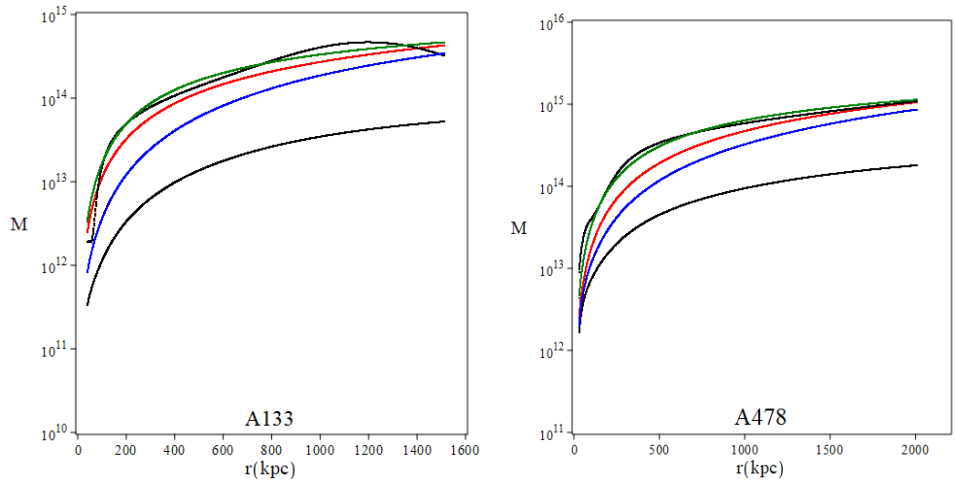

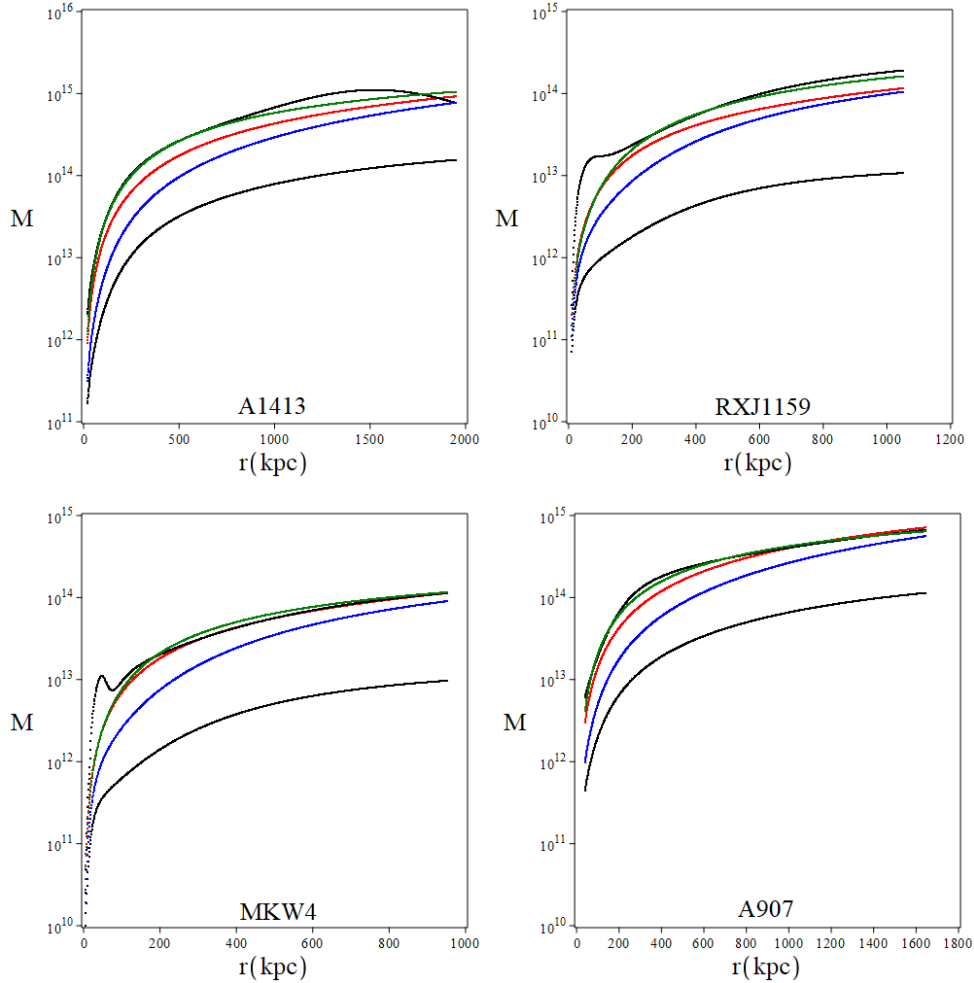

**Figure 4:** Mass profiles in units of M$_\odot$: M$_{bar}$ (lower black line), M$_{dyn}$ (upper black line), M$_{DM}$ (green line), original GRAS/AQUAL field equation (35) (blue line), modified GRAS/AQUAL field equation (44) (red line)

As expected, the new GRAS/AQUAL field equation (44) leads to much better agreement between M$_{GA}$ and the M$_{dyn}$, M$_{DM}$ profiles, than the current GRAS/AQUAL field equation (35) does. Although the M$_{GA}$ profiles still appear to fall somewhat short of M$_{dyn}$, it is reasonable to take the field equation given by Eq. (44) as an approximation to the true field equation, whatever that turns out to be.

On Figure 5 are shown the 6 clusters plotted on the RAR plot of Figure 1. Although each cluster, with their individual baryonic mass distribution, will behave differently, there are some common trends. For $\log(g_{bar}) > -11$, the cluster values tend towards the cluster RAR, while for $\log(g_{bar}) < -11$, the values tend towards the galactic RAR. This is expected as one travels from inside the baryonic mass distribution towards the outside of the baryonic mass distribution. Also plotted on the figure are two sets of the cluster data from Tian et al (2020). These correspond to the outer most data values for their cluster samples, specifically at r = 400 kpc and 600 kpc. Although the data is limited, there is a hint of a trend in line with the theoretical behaviour found. Ideally future measurements can be expanded well beyond 600 kpc to see if this trend is real. This will have significant consequences as will be discussed in the conclusion.

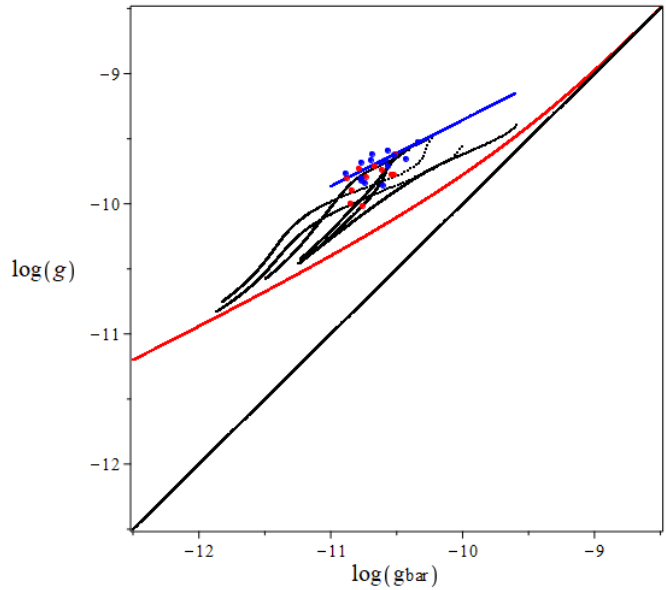

**Figure 5:** The dependence that g has on $g_{bar}$ for the 6 cluster samples (black lines) compared to the galactic RAR (red line), and the cluster RAR (blue line). Included are the data points from Tian et al (2020) corresponding to cluster radial distances of 400 kpc (blue data points), and 600 kpc (red data points).

**4.2 Galaxies**

Substituting $r_o$ = 150 kpc and $g_{GAo}$, as given by Eq. (45), into Eq. (44) allows for the gravitational field profile, $g_{GA}$, to be determined for each of the 3 galaxies. The resulting velocity curves for the 3 galaxies are shown in Figure 6. Included in the figures are the velocity curves calculated from the original GRAS/AQUAL field equation and the ones that fall out directly from the galactic RAR, Eq. (2).

There certainly are differences between the velocity curves determined from the modified field equation and the RAR velocity curves. These differences are due to the parameter β not being exactly equal to 1. As a comparison, the velocity curves derived from the original

GRAS/AQUAL field equation do correspond to β =1. However, it must be kept in mind that the interpolating function was chosen to provide a good fit between the original GRAS/AQUAL field equation and the RAR. Overall, the difference found is not great enough to disqualify the modified GRAS/AQUAL field equation as an approximation to the true field equation. On Figure 7 are shown values from the 3 modeled galaxies along with the galaxy and cluster RAR's. Again, the differences found between the values that fall out of the modified field equation and the galactic RAR are relatively small.

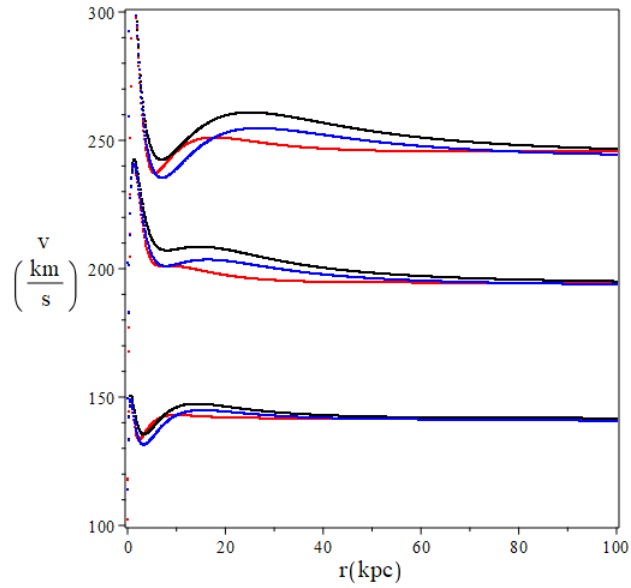

**Figure 6:** Rotational velocity curves for the modeled galaxies. Generated from the galactic RAR equation (2) (black line), generated from the original GRAS/AQUAL equation (35) (blue line), generated from the modified GRAS/AQUAL equation (44) (red line).

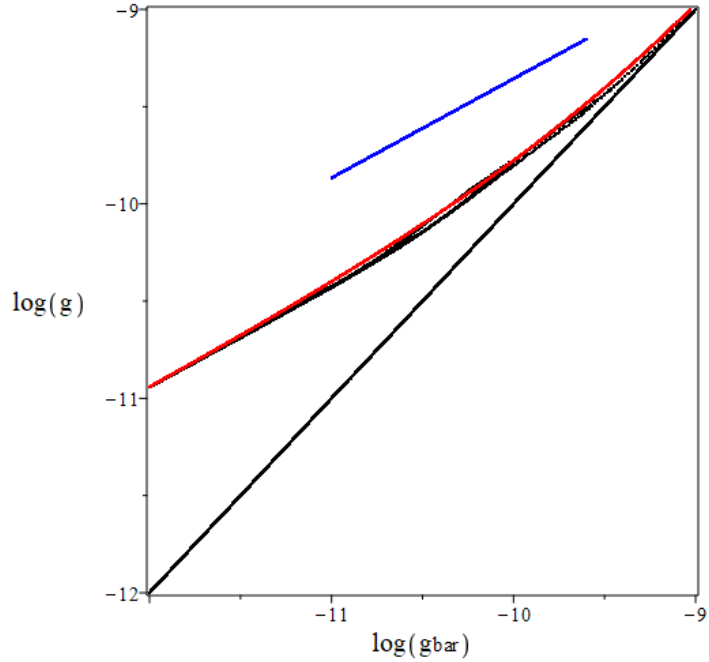

**Figure 7:** The dependence that g has on $g_{bar}$ for the 3 modeled galaxies (black lines) compared to the galactic RAR (red line) and the cluster RAR (blue line).

Ideally, comparisons should be made with real galaxies, and the effect of different interpolation functions should be considered. However, the simple galaxy model used does allow for the general effect of the modified field equation on clusters and galaxies to be compared.

**4.3 Solar System**

In addition to checking the effect that the modified GRAS/AQUAL field equation has on galaxies, the possible effect on solar system dynamics needs to be considered. Using positional observations of planets and spacecraft, Pitjev and Pitjeva (2013) have determined that the residual precession rate for Saturn is $-(0.32 \pm 0.47)$ mas cy$^{-1}$ and that any unaccounted-for mass, $\Delta M$, within Saturn's orbit (~10 AU) must be $< 1.7 \times 10^{-10}$ M$_\odot$.

Within the solar system, the gravitational field falls off as $r^{-2}$, and therefore $\beta$ is equal to ½. The field equation as given by Eq. (41) then becomes

$$g_{GA} = g_{bar} + \frac{1}{2} f(g_{GA}/g_o) g_{GA}. \tag{47}$$

Solving Eq. (47) for r = 10 AU, leads to $g_{GA}(10\ AU) - g_{bar}(10\ AU) = 4.88 \times 10^{-16}\ ms^{-2}$, corresponding to an additional mass within Saturn's orbit of $8.28 \times 10^{-12}\ M_\odot$. This is well within Pitjev and Pitjeva's limit.

## 5. CONCLUSION

A modified GRAS/AQUAL field equation was developed and applied to a sample of six galaxy clusters. These six clusters were chosen as their given baryonic mass profiles were found to approach a constant value at large distances, allowing them to be treated as being bounded. This modified field equation contains just a single parameter, namely $g_o$, the same parameter found with the galactic RAR. However, the modified field equation now involves both the derivative of the gravitational field as well as the gravitational field itself. In terms of GRAS, this corresponds to the mass dipole moment density depending on both the gravitational field and its derivative.

The new field equation leads to the gravitational field depending on the complete baryonic mass distribution and not just on the total baryonic mass within the selected radius. This dependence allows the modified field equation to distinguish between galaxies and galaxy clusters, which have very distinctive baryonic mass distributions.

The resulting total mass profiles, determined by using the new field equation, were found to be in reasonably good agreement with the total mass profiles based on the stability of the intracluster gas. The modified field equation also led to velocity curves for a set of modeled galaxy baryonic mass distributions that were in fair agreement with those obtained from the galactic RAR and the current GRAS/AQUAL field equation. Overall, it is concluded that a single field equation that involves just a single free parameter can account for the missing mass found with both galaxies and clusters. The particular field equation provided in this manuscript should be looked upon as being an approximation to the actual field equation, whatever that may be.

The dependence that the modified GRAS/AQUAL field equation has on the baryonic mass profile also leads to an understanding of the two RAR's, Eq. (2) for galaxies, and Eq. (7) for galaxy clusters. The galactic RAR is applicable to both inside and outside of a galaxies baryonic mass distribution, while the cluster RAR is only applicable within the intracluster gas region of clusters. Crucially, far outside of a clusters baryonic mass distribution the galactic RAR should also hold. If there is a universal gravitational field equation, this must be the case. Far outside of their baryonic mass distributions, galaxies and clusters must behave similarly. This leads directly to a method of determining if the additional gravitational fields of galaxies and clusters are due not to dark matter but to incomplete gravitational theory. If measurements can be taken out to the further most regions of clusters, beyond the current limit of 600 kpc of Tian et al (2020), the question is, does the galactic RAR given by Eq. (2) begin to apply? If it does, then alternate gravitational theories such as MOND and GRAS are on the right track. If not, dark matter would be the most probable answer.

**Funding:**    none

**Ethical Statement:**   none

**Author contribution:**    A. Raymond Penner is the sole author

**Data availability statement:** none

**Informed consent:**   none

**REFERENCES**

Aghanim N., Akrami Y., Arroja F., 2020 et al., A&A, 641, A1

Angus G.W., Shan H.Y., Zhao H.S., Famaey B., 2007, ApJ, 654, L13

Aurelien H., Benoit F., Angus G.W., Gentile G., 2016, MNRAS, 455-1, 449

Banik I. & Zhao H., 2018, ScieFed Journal of Astrophysics, 1, 1000008

Banik I. & Zhao H., 2022, Symmetry, 14- 7, 1331

Banik I. & Kroupa P., 2020, MNRAS, 495-4, 3974

Bekenstein J.D. & Milgrom M., 1984, ApJ, 286, 7

Bertone G. & Hooper D., 2018, Rev. Mod. Phys., 90, 045002

Bilek M., Muller O., Famaey B., 2019, A&A, 627, L1

Blanchard A., Heloret J-Y., Ilic S., Lamina B., Tutusaus I., 2022, arXiv:2205.05017

Blanchet L., 2007a, Class. Quant. Grav., 24, 3529, astro-ph/0605637

Blanchet L. 2007b, Class. Quant. Grav., 24, 3541, gr-qc/0609121

Blanchet L. & LeTiec A., 2008, arXiv.org/abs/0804.3518

Cesare V., Diaferio A., Matsakos T., & Angus G., 2020, A&A, 637

De-Chang D, Matsuo R., Starkman G., 2010, Phys. Rev. D, 81(2)

Desmond H., 2017a, MNRAS, 464-4, 4160

Desmond, H. 2017b, MNRAS, 472-1, L35

Famaey B. & McGaugh S., 2012, Living Reviews in Relativity, 15, 10

Famaey B., Khoury J., Penco R., 2018, J. Cosmol. Astropart. Phys. 1803, 038

Hajdukovic D.S., 2011a, Astrophys. Space Sci., 334, 215

Hajdukovic D.S., 2011b, Adv. Astron., 196, 852

Hajdukovic D.S., 2020, MNRAS, 491-4, 4816

Hees A., Folkner W. M., Jacobson R. A., Park R. S., 2014, Phys. Rev. D, 89, 102002

Hees A., Famaey B., Angus G.W., Gentile G., 2016, MNRAS, 455, 449

Hodson A.O. & Zhao H., 2017a, A&A, 607, A109

Hodson A.O. & Zhao H., 2017b, arXiv:1701.03369v1

Koda J., Yagi M., Yamanoi H. & Komiyama Y., 2015, ApJ, 807, L2

Knebe A., Llinares C., Wu X., Zhao, H., 2009, ApJ, 703, 2285

Kroupa P., 2014, Can. J. Phys., 93-2

Kroupa P., Banik I, Haghi H., et al, 2018, Nat. Astron. 2, 927

Lelli F., McGaugh S.S., Schombert J.M., Pawlowski M.S., 2017, ApJ, 836-2, 152

Ludlow A.D., Benitez-Llambay A., Schaller M., et al, 2017, Phys. Rev. Lett. 118, 161103

Ma L., Zhang Z., Wang H, Wu. X., 2025, ApJ, 948, 101

Matsakos T., & Diaferio A. 2016, ArXiv e-prints [arXiv:1603.04943]

McGaugh S.S., de Blok W.J.G., 1998, arXiv:Astro-ph/9801102v1

McGaugh S.S., Lelli F., & Schombert J.M., 2016, Phys. Rev. Lett., 117, 201101

McGaugh S. S., Schombert J. M., Bothun G. D., de Blok W. J. G., 2000, Astrophysical Journal Letters, 533-2, L99

McGaugh S.S., 2012, Astron. J, 143, 40

McGaugh S.S., 2015, Can. J. Phys., 93(2), 250

McGaugh S.S., Li P., Lelli F., Schombert J.M., 2018, Nat. Astron. 2, 924

Mihos J.C., Durrell P.R., Ferrarese L., et al, 2015, ApJ, 809, L21

Milgrom M., 1983a, ApJ, 270, 365

Milgrom M., 1983b, ApJ, 270, 371

Milgrom M., 1983c, ApJ, 270, 384

Milgrom M., 1986, ApJ, 302, 617

Mistele T., McGaugh S., Lelli F., Schombert J., Li, P., 2024, Astrophysical Journal Letters, 969, L3

Navarro J.F., Frenk C.S., White S.D.M., 1996, ApJ, 462, 563

Navarro J.F., Frenk C.S., White S.D.M., 1997, ApJ, 490, 493

Navarro J.F., Benitez-Llambay A., Fattahi A., et al, 2017, MNRAS, 471, 1841

Pawlowski M.S., 2021, Nature Astronomy, 5, 1185

Penner A.R., 2011, Can. J. Phys., 89, 841

Penner A.R., 2013, Can. J. Phys., 91, 610

Penner A.R., 2016a, Astrophys. Space Sci., 361, 124

Penner A.R., 2016b, Astrophys. Space Sci., 361, 361

Penner A.R., 2018, Gravitational Anti-Screening as an Alternative to Dark Matter, in Research Advances in Astronomy, ed. N. Mehler. (NY, NY: Nova US), 1

Penner A.R., 2020a, Astrophys. Space Sci., 365, 65

Penner A.R., 2020b, Astrophys. Space Sci., 365, 154

Penner A.R., 2022, Class. Quantum Grav., 39

Penner A.R., 2023, MNRAS, 522-3, 4003

Pitjev N.P., Pitjeva E.V., 2013, Astron. Lett. 39(3), 141

Read J.I., Trentham N., 2005, Phil. Trans : Math. Phys. & Eng. Sci., 363, 1837

Rodrigues D.C., Marra V., Del Popolo A., Davari Z., 2018a, Nat. Astron. 2, 668

Rodrigues D.C., Marra V., Del Popolo A., Davari Z., 2018b, Nat. Astron. 2, 927

Rodrigues D.C., Marra V., Del Popolo A., Davari Z., 2020, Nat. Astron. 4, 134

Roshan M., Ghafourian N., Kashi T., Banik I., Haslbauer M., Cuomo V., Famaey B., Kroupa P., 2021, arXiv:2106.10304v3

Sanders R.H., 2003, MNRAS, 342-3, 901

Sanders R.H., 2014, Can. J. Phys., 93-2, 126

Tian Y., Umetsu K.,Ko C.-M., Donahue M., & Chiu I.-N, 2020, ApJ, 896, 70

Tian Y., Ko C.-M., Li P., McGaugh S., & Poblete S.L., 2024, A&A, 684, A180

Trimble V., 1987, Ann. Rev. Ast. Astrophys., 25(1)

Tully R. B., Fisher J. R., 1977, A&A, 54, 661

Vikhlinin A., Kravtsov A., Forman W., et al, 2006, ApJ, 640, 691

Wu X., Kroupa P., 2015, MNRAS, 446-1, 330

Zhao H. & Famaey B., 2012, Phys. Rev. D., 86, 067301

Zwicky F., 1933, Helv. Phys. Acta., 6, 110